\documentstyle[multicol,epsf,aps]{revtex}
\begin{document}
\title{Random Fibonacci Sequences}
\author{Cl\'ement Sire$^1$ and Paul L. Krapivsky$^2$}
\address{$^1$Laboratoire de Physique Quantique (UMR 5626 du CNRS),
Universit\'e Paul Sabatier, 31062 Toulouse Cedex, France\\
$^2$Center for Polymer Physics and Department of Physics, Boston
University, Boston MA 02215, USA}
\maketitle

\begin{abstract}
  Solutions to the random Fibonacci recurrence $x_{n+1}=x_n \pm \beta
  x_{n-1}$ decrease (increase) exponentially, $x_n\sim \exp(\lambda n)$, for
  sufficiently small (large) $\beta$.  In the limits $\beta\to 0$ and
  $\beta\to\infty$, we expand the Lyapunov exponent $\lambda(\beta)$ in
  powers of $\beta$ and $\beta^{-1}$, respectively.  For the classical case
  of $\beta=1$ we obtain exact non-perturbative results.  In particular, an
  invariant measure associated with Ricatti variable $r_n=x_{n+1}/x_n$ is
  shown to exhibit plateaux around all rational $r$'s.

\smallskip\noindent{PACS numbers: 02.50.-r, 31.15.Md, 72.15.Rn}
\end{abstract}

\begin{multicols}{2}

\section{Introduction}

The Fibonacci numbers $1,1,2,3,5,8,13,\ldots$ defined via
$F_{n+1}=F_n+F_{n-1}$ abound in nature\cite{conway,http}. For example, they
govern the number of leaves, petals and seed grains in plants\cite{J}; they
also give the number of ancestors of a drone.  The Fibonacci recurrence is
the simplest recurrence in which each number depends on the previous two and
this perhaps explains why the Fibonacci sequence is so ubiquitous. A natural
{\em stochastic} modification of the Fibonacci sequence is to allow both
additions and substractions.  Random Fibonacci sequences are related to many
fields including condensed matter physics, dynamical systems, products of
random matrices\cite{BL}, continued fractions, etc.  Random recurrences also
form a chapter of the larger subject of iterated random functions\cite{DF}.

The random Fibonacci recurrence $x_{n+1}=x_n \pm x_{n-1}$ results in
sequences which behave erratically for small $n$.  In the limit $n\to\infty$,
however, exponential growth occurs with probability one as was established by
Furstenberg\cite{F} in 1963.  The large $n$ behavior is characterized by the
Lyapunov exponent $\lambda$,
\begin{equation}
\label{Le}
\lambda=\lim_{n\to\infty} {\ln |x_n|\over n}.
\end{equation}
Exponential growth (decay) means that $\lambda$ is positive (negative).  For
the random Fibonacci recurrence where each $\pm$ sign is independent and
either $+$ or $-$ with probability 1/2, the Lyapunov exponent is
$\lambda=0.12397559\ldots$\cite{V}.  Exponential growth may seem unsurprising
but similar generalized random Fibonacci recurrence,
\begin{equation}
\label{rFr}
x_{n+1}=x_n \pm \beta x_{n-1},
\end{equation}
gives exponential growth only when the parameter $\beta$ is sufficiently
large, $\beta>\beta_s\approx 0.70258$, whereas for $0<\beta<\beta_s$
solutions decay exponentially\cite{E+T}.  The decay occurs even though the
expected values $\langle x_n\rangle$ remain constant and the expected values
of the higher integer moments $\langle x_n^2\rangle$, $\langle x_n^3\rangle$,
etc.  grow exponentially.  Indeed, from Eq.~(\ref{rFr}) one can deduce
$\langle x_{n+1}^2\rangle=\langle x_n^2\rangle +\beta^2 \langle
x_{n-1}^2\rangle$ which implies $\langle x_n^2\rangle\sim \left[{1\over
    2}+({1\over 4}+\beta^2)^{1/2}\right]^n$, and similarly for higher
moments\cite{gleb}.  Additionally, for $\beta>1/4$ an interesting non-smooth
dependence of the Lyapunov exponent on the parameter $\beta$ has been
observed\cite{E+T} suggesting that the curve $\lambda(\beta)$ is a fractal
(it remains unclear whether this curve becomes genuinely smooth for
sufficiently large $\beta$).  Similar non-smooth dependence on parameters has
been reported for several disordered systems\cite{BH,D+H}.  Also, numerical
results\cite{E+T} suggest the following asymptotic behaviors of the Lyapunov
exponent:
\begin{equation}
\label{finds}
\lambda\to \cases{
-{1\over 2}\,\beta^2 -{15\over 8}\,\beta^4 &when $\beta\to 0$,\cr
\cr
{1\over 2}\,\ln\beta+{0.114\over\beta} &when $\beta\to \infty$.\cr}
\end{equation}

While understanding of the nature of the curve $\lambda(\beta)$ might be a
very difficult problem, one should be able to carry out asymptotic expansions
of the Lyapunov exponent using tools developed in studies of one-dimensional
disordered systems\cite{BH,D+H,D,S,K+M,D+G,CPV,JML}.  Indeed, Eq.~(\ref{rFr})
admits the standard interpretation in terms of the one-dimensional
(discretized) Schr\"odinger equation.  A peculiarity of the present problem
is that the corresponding random Hamiltonian is non-Hermitian.  Non-Hermitian
random Hamiltonians appear in various non-equilibrium problems\cite{noneq}
and exhibit interesting behaviors\cite{H+N,G+K,F+Z,DJZ}, e.g., a
delocalization transition may occur even in one dimension.  Fortunately,
tools developed for Hermitian problems can often be extended to the
non-Hermitian case.

In the next Sect.~II, we employ perturbation theory to expand the Lyapunov
exponent in powers of $\beta$. In particular, we show that in the $\beta\to
0$ limit the second term is $-1.75\times \beta^4$ rather than $-1.875\times
\beta^4$ [cf.  Eq.~(\ref{finds})] and compute many more terms.  In Sec.~III,
we use non-perturbative techniques to analyze the special $\beta=1$ case.  In
Sec.~IV, we examine random sequences with Gaussian, rather than binary,
disorder.  Finally, a few open questions are discussed in the last Sec.~V.

\section{Perturbation Theory}

An old way of studying linear random recursions is to reduce them to random
maps by introducing the Ricatti variable $R_n=x_{n+1}/x_n$.  In the present
case, Eq.~(\ref{rFr}) becomes
\begin{equation}
\label{Rn}
R_n=1 \pm {\beta\over R_{n-1}}.
\end{equation}
The Lyapunov exponent is given by
\begin{equation}
\label{Lexp}
\lambda=\lim_{n\to\infty} {1\over n}\,\sum_{j=1}^n \ln R_j.
\end{equation}
In the large $n$ limit, the probability distribution of $R_n$ becomes
independent on $n$.  The invariant distribution $P(R,\beta)$ satisfies
\begin{equation}
\label{PRfun}
P(R)={\beta\over 2(R-1)^2}\left[P\left({\beta\over R-1}\right)+
P\left({\beta\over 1-R}\right)\right]
\end{equation}
which immediately follows from the random map (\ref{Rn}).  The analysis of
the functional equation (\ref{PRfun}) often simplifies after transforming it
into an integral equation
\begin{eqnarray}
\label{PR}
P(R,\beta)=&&\int dR'\,P(R',\beta)\\
&&\times{1\over 2}\left[\delta\left(R-1+{\beta\over R'}\right)
+\delta\left(R-1-{\beta\over R'}\right)\right].\nonumber
\end{eqnarray}
Once we know the invariant distribution $P(R,\beta)$, we can employ
Eq.~(\ref{Lexp}) to compute the Lyapunov exponent
\begin{equation}
\label{Lb}
\lambda(\beta)=\int dR\,P(R,\beta)\,\ln R.
\end{equation}

Depending on the magnitude of $\beta$, the support of the invariant
distribution $P(R,\beta)$ may be either finite or infinite.  Let us assume
the former.  Then the extreme values satisfy $R_{\rm min}= 1-\beta/R_{\rm
  min}$ and $R_{\rm max}=1+\beta/R_{\rm min}$, or
\begin{eqnarray}
\label{Rmm}
R_{\rm min}={1+\sqrt{1-4\beta}\over 2},\quad
R_{\rm max}={3-\sqrt{1-4\beta}\over 2}.
\end{eqnarray}
Thus, the support is finite when the strength of disorder satisfies
$\beta<1/4$.  Furthermore, in this case the support is not merely the
interval $[R_{\rm min},R_{\rm max}]$ but rather a fractal set similar to the
Cantor set.  Indeed, the map $R\to 1\pm\beta/R$ transforms the interval
$[R_{\rm min},R_{\rm max}]$ into the union of two subintervals, $[R_{\rm
  min},1-\beta/R_{\rm max}]$ and $[1+\beta/R_{\rm max}, R_{\rm max}]$.
Restricting the map to these two subintervals shows that they are
transformed into four subintervals.  Proceeding in the same
manner {\em ad infinitum} we construct the fractal support of the
invariant distribution.  On the other hand, the support is
unbounded when $\beta>1/4$. This suggests to employ different
perturbation approaches for small and large $\beta$.

\subsection{Weak disorder expansion}

When $\beta<1/4$, it is desirable to compute $\lambda(\beta)$ without
explicit determination of the invariant distribution $P(R,\beta)$ which is a
complicated singular function.  The trick is to transform the integral in the
basic relation (\ref{Lb}) into an integral which can be calculated
perturbatively using only the normalization requirement $\int dR\,P(R)=1$.
To this end we insert Eq.~(\ref{PR}) into Eq.~(\ref{Lb}) to yield
\begin{equation}
\label{Lb1}
\lambda={1\over 2}\int dR\,P(R)
\left[\ln\left(1-{\beta\over R}\right)
+\ln\left(1+{\beta\over R}\right)\right].
\end{equation}
Expanding the logarithms on the right-hand side of Eq.~(\ref{Lb1}) we obtain
\begin{eqnarray*}
\lambda=-{\beta^2\over 2}\int dR\,R^{-2}P(R)
-{\beta^4\over 4}\int dR\,R^{-4}P(R)+\ldots
\end{eqnarray*}
In the limit $\beta\to 0$, the interval $[R_{\rm min},R_{\rm max}]$ shrinks
to $R=1$.  Hence, the first integral on the right-hand side of the above
equation approaches to $\int dR\,P(R)=1$.  Thus $\lambda=-\beta^2/2+{\cal
  O}(\beta^4)$ and the first term of the expansion was indeed derived without
using an explicit form of the invariant distribution.  However, the
derivation of the next term is still impossible without knowledge of the
invariant distribution.  To avoid this we transform Eq.~(\ref{Lb1}) as we
transformed Eq.~(\ref{Lb}) into Eq.~(\ref{Lb1}).  Namely, we plug
Eq.~(\ref{PR}) into Eq.~(\ref{Lb1}) to give
\begin{equation}
\label{Lb2}
\lambda={1\over 4}\int dR\,P(R,\beta)\,L_2(R,\beta)
\end{equation}
where
\begin{eqnarray*}
L_2(R,\beta)=&&\ln\left(1-{\beta\over 1-{\beta\over R}}\right)
               +\ln\left(1-{\beta\over 1+{\beta\over R}}\right)\\
             &&+\ln\left(1+{\beta\over 1+{\beta\over R}}\right)
               +\ln\left(1+{\beta\over 1-{\beta\over R}}\right).
\end{eqnarray*}
Expanding the logarithms and inserting this expansion into Eq.~(\ref{Lb2}) we
obtain
\begin{eqnarray*}
\lambda=-{\beta^2\over 2}\int dR\,P(R)
-{\beta^4\over 4}\int dR\,\left({6\over R^2}+1\right)P(R)+\ldots
\end{eqnarray*}
The first integral is now identically equal to one, while the second
approaches to 7 in the small $\beta$ limit.  Therefore, $\lambda=-\beta^2/2
-7\beta^4/4+{\cal O}(\beta^6)$ is the two-term expansion.  This shows that
the small $\beta$ expansion of Eq.~(\ref{finds}) is slightly incorrect.

Repeating the above procedure $k$ times, we recast Eq.~(\ref{Lb}) into
\begin{equation}
\label{Lbk}
\lambda=2^{-k}\int dR\,P(R,\beta)\,L_k(R,\beta),
\end{equation}
where
\begin{equation}
\label{Lk}
L_k(R,\beta)=\sum_{\epsilon_1,\ldots,\epsilon_k}
\ln\,[1;\epsilon_1 \beta, 1, \epsilon_2 \beta,\ldots, \epsilon_k \beta, R].
\end{equation}
The sum on the right-hand side of Eq.~(\ref{Lk}) is taken over all
$\epsilon_j=\pm 1$, and $[1; a, b,\ldots]$ denotes the continued fraction
$1+{a\over b+\ldots}$.  Expanding $L_k(R,\beta)$ and plugging it into
Eq.~(\ref{Lbk}) one finds the correct expansion of the Lyapunov exponent up
to ${\cal O}(\beta^{2k})$.  A few of these terms can be computed by hand, and
a bit more can be done with the help of Mathematica. One gets
\begin{eqnarray}
\label{smallbeta} \lambda(\beta)=&&-{1\over 2}\,\beta^2 -{7\over
4}\,\beta^4
-{29\over 3}\,\beta^6-{555\over 8}\,\beta^8\nonumber\\
&&-{2843\over 5}\,\beta^{10}
-{30755\over 6}\,\beta^{12}+{\cal O}(\beta^{14}).
\end{eqnarray}
The radius of convergence of this series appears to be equal to 1/4 as one
could guess from Eqs.~(\ref{Rmm}).  Hence the Lyapunov exponent is an
analytic function of $\beta$ when $|\beta|<1/4$. Amusingly, the invariant
distribution is a very singular function in this range of the parameter
$\beta$. The series (\ref{smallbeta}) perfectly reproduces numerical
results\cite{E+T} (5 representative digits), except for the case $\beta=1/4$
for which $\lambda_{\rm num}\approx -0.0429$ and $\lambda_{\rm theor}\approx
-0.0424$. This slight discrepancy is due to the fact that for $\beta=1/4$ the
generic term of the series decays algebraically in contrast with an
exponential decay for $\beta<1/4$.

\subsection{Strong disorder expansion}

In the large $\beta$ limit, the support of the invariant distribution
$P(R,\beta)$ is the whole real line.  The trick which has been employed in
the case of weak disorder does not apply, i.e., we cannot determine
$\lambda(\beta)$ without knowledge of the invariant distribution.  It proves
convenient to use the modified Ricatti variable
$r_n=|x_{n+1}/x_n\sqrt{\beta}|$.  Then, Eq.~(\ref{rFr}) reduces to the random
map
\begin{equation}
\label{Rnbis}
r_n=\left|{1\over r_{n-1}}\pm \delta\right|, \qquad
\delta\equiv \beta^{-1/2}.
\end{equation}
Once we know the invariant distribution $P(r,\delta)$, we can compute the
Lyapunov exponent via Eq.~(\ref{Lb}) which now becomes
\begin{equation}
\label{lb}
\lambda={1\over 2}\,\ln \beta+\int dr\,P(r)\,\ln r.
\end{equation}
The support of the invariant distribution is $0\leq r<\infty$. It proves
convenient to define $P(r)$ for negative $r$ via $P(r)=P(-r)$, so the support
is the entire real line.  In present notations, the functional equation
(\ref{PRfun}) becomes
\begin{equation}
\label{Prdif}
2P(r)={1\over (r+\delta)^2}P\left({1\over r+\delta}\right)\!+\!
{1\over (r-\delta)^2}P\left({1\over r-\delta}\right).
\end{equation}
For $\delta=0$, this equation reduces to $P(r)=r^{-2}P(1/r)$ which has
infinitely many solutions. For $\delta>0$, however, the invariant
distribution is unique.  Thus taking the $\delta\to 0$ limit of the invariant
distribution $P(r,\delta)$ we shall obtain a unique appropriate solution. In
this sense, we are building a degenerate perturbation approach.

To determine $P(r,\delta)$, notice that Eq.~(\ref{Prdif}) can be re-written
as
\begin{equation}
\label{Pr}
P(r)=r^{-2}P(1/r)+{1\over 2}\,D^2_\delta \left[r^{-2}P(1/r)\right],
\end{equation}
where $D^2_\delta$ is the discrete analog of the second derivative,
$D^2_\delta F(r)=F(r+\delta)-2F(r)+F(r-\delta)$.  Expanding the right-hand
side of Eq.~(\ref{Pr}) into Taylor series and denoting $D={d\over dr}$ we
obtain
\begin{equation}
\label{P1}
P(r)=\sum_{k=0}^\infty {\delta^{2k}\over (2k)!}\,
D^{2k}\left[r^{-2}P(1/r)\right].
\end{equation}
The change of variable $r\to 1/r$ transforms Eq.~(\ref{P1}) into
\begin{equation}
\label{P2}
P(1/r)=\sum_{k=0}^\infty {\delta^{2k}\over (2k)!}\,
{\cal D}^{2k}\left[r^2 P(r)\right],
\end{equation}
where ${\cal D}=r^2\,{d\over dr}$.  If we now plug Eq.~(\ref{P2}) into
Eq.~(\ref{P1}), we eliminate $P(1/r)$ and thus arrive at a closed equation
for the invariant distribution $P(r,\delta)$:
\begin{equation}
\label{P}
P(r)=\sum_{k=0}^\infty \sum_{l=0}^\infty {\delta^{2k+2l}\over (2k)!(2l)!}\,
D^{2k}\left\{r^{-2}{\cal D}^{2l}\left[r^2 P(r)\right]\right\}.
\end{equation}
Equation (\ref{P}) suggests to seek a solution $P(r,\delta)$ which can be
expanded as a power series in $\delta^2$:
\begin{equation}
\label{Pser}
P(r,\delta)=\sum_{j=0}^\infty \delta^{2j} P_j(r).
\end{equation}
Plugging the series (\ref{Pser}) into Eq.~(\ref{P}) one finds that the terms
of order one cancel.  Equating terms of order $\delta^2$ leads to the
following second order differential equation for $P_0(r)$
\begin{equation}
\label{P02}
D{\cal L}_1\,P_0(r)=0 \quad {\rm with}\quad {\cal L}_1\equiv D+r^{2}Dr^{2}.
\end{equation}
Integrating once we arrive at ${\cal L}_1P_0=0$, the integration constant
being equal to zero to ensure that $P_0(r)$ vanishes in the large $r$ limit.
This can be re-written as
\begin{equation}
\label{Pr0diff}
(1+r^4)P_0'(r)+2r^3P_0(r)=0,
\end{equation}
whose normalized solution is
\begin{equation}
\label{P0}
P_0(r)={2\sqrt{\pi}\over \Gamma^2(1/4)}\,{1\over \sqrt{1+r^4}}.
\end{equation}
Thus among infinitely many invariant distributions satisfying the
duality relation $P(r)=r^{-2}P(1/r)$ we have indeed selected the
specific solution (\ref{P0}).  Note that $\int dr\,P_0(r)\ln r=0$
(this is actually valid for any function which obeys the duality
relation).  Therefore, the anticipated contribution of order one
to the Lyapunov exponent [cf. Eq.~(\ref{lb})] is actually equal
to zero.

Similarly equating the terms of order $\delta^4$ one finds $D{\cal
  L}_1P_1+D{\cal L}_3P_0=0$, where ${\cal L}_3$ is certain third order
differential operator.  Integrating the above equation yields ${\cal
  L}_1P_1+{\cal L}_3P_0=0$ (the integration constant is equal to zero to
ensure that $P_1(r)$ vanishes in the large $r$ limit).  Recalling that ${\cal
  L}_1P_0=0$ one can simplify the term ${\cal L}_3P_0$.  The final first
order differential equation for $P_1(r)$ reads
\begin{equation}
\label{P12}
{\cal L}_1P_1(r)={(5+r^4)D^3+2r^{3} D^2\over 12}\,P_0(r).
\end{equation}
It is helpful to write $P_1(r)=P_0(r)\,f_1(r)$ and then exploit the identity
${\cal L}_1 [P_0(r)f_1(r)]\equiv (1+r^4)P_0(r)f_1'(r)$.  Integrating the
resulting equation yields
\begin{eqnarray}
\label{r1}
f_1(r)=c_1+R_1(r), \quad
R_1(r)=\frac{r^2(3r^8+8r^4-15)}{6(r^4+1)^3}.
\end{eqnarray}
The integration constant $c_1$ can be found from the relation $\int
dr\,P_1(r)=0$ which ensures that the normalization condition $\int
dr\,P(r)=1$ holds.  We get
\begin{eqnarray}
c_1=\left[{\Gamma(3/4)\over \Gamma(1/4)}\right]^2.
\end{eqnarray}
A direct computation gives $\int dr\,P_1(r)\ln r=c_1$.  Thus, in this order
the contribution to the Lyapunov exponent is equal to $c_1\beta^{-1}$.

To determine $P_j(r)$, we repeat the above procedure.  Plugging (\ref{Pser})
into (\ref{P}), equating terms of order $\delta^{2j+2}$, and integrating once
the resulting equation yields
\begin{equation}
\label{Pj}
{\cal L}_1P_j(r)+\sum_{n=1}^{j} {\cal L}_{2n+1}P_{j-n}(r)=0,
\end{equation}
where ${\cal L}_{2n+1}$ is the differential operator of order $2n+1$,
\begin{eqnarray*}
{\cal L}_{2n+1}&=&{2\over (2n+2)!} D^{2n+1}\\
&+&\sum_{k=0}^n {2\over (2k)!(2n-2k+2)!}D^{2k}{\cal D}^{2(n-k)+1}r^2.
\end{eqnarray*}
Equations (\ref{Pj}) can be solved recursively.  Writing again
$P_j(r)=f_j(r)P_0(r)$, we find in the second order
\begin{eqnarray*}
f_2(r)&=&-{1\over 96}+c_1f_1(r)+R_2(r),\\
R_2(r)&=&\frac{r^4(463-3640r^4+2514r^{8}+440r^{12}-
17r^{16})}{24(r^4+1)^{6}}.
\end{eqnarray*}
In the third order we have
\begin{eqnarray*}
f_3(r)&=&{14699\over 21600}\,c_1-{1\over 96}\,f_1(r)+c_1f_2(r)+R_3(r)\\
R_3(r)&=&{r^2\,Q_3(r)\over 15120(r^4+1)^9}\\
Q_3(r)&=&11340r^{32}-678825r^{28}-11260368r^{24}\\
&-&3619377r^{20}+356871272r^{16}-471736467r^{12}\\
&+&125696592r^8-5587155r^4+11340.
\end{eqnarray*}
The constants were determined recursively after a long computation exploiting
the normalization requirements $\int dr\,P_j(r)=0$ and the basic identity
$\Gamma(1+z)=z\Gamma(z)$ for the gamma function\cite{BO}.

After inserting above expressions for $P_j(r)$ in Eq.~(\ref{lb}), we arrive
at the expansion
\begin{eqnarray}
\label{lbdev} \lambda(\beta)={1\over 2}\ln\beta+\sum_{k=1}^\infty
a_k\beta^{-k}.
\end{eqnarray}
One can in principle generate $\lambda(\beta)$ to any order.  The first few
coefficients are
\begin{eqnarray*}
&&a_1=c_1=0.1142366452611159\ldots\\
&&a_2=c_1^2+{1\over 144}=0.01999445564958\ldots\\
&&a_3=0.0105345239\ldots\\
&&a_4=0.0176632096\ldots
\end{eqnarray*}
The coefficients $a_i$'s have extremely complicated analytical expressions in
terms of the $\Gamma$ function which we have not been able to simplify
(although it appears plausible that these expressions can be reduced to
polynomials of $c_1$ with rational coefficients).  The exact value for $a_1$
is in good agreement with the numerical estimate from Ref.\cite{E+T}.  {}From
the first four coefficients, one could guess that the radius of the
convergence of series (\ref{lbdev}) is of order unity.  However, an apparent
fractal structure of the curve $\lambda(\beta)$ makes the above guess
questionable.  Even if our strong disorder expansion is asymptotic, it is
quite accurate as can be seen by comparison of the four-term expansion
(\ref{lbdev}) with numerical results.  For $\beta=8$, both give
$\lambda\approx 1.05433$ whereas there is a slight discrepancy for $\beta=4$
as $\lambda_{\rm num}\approx 0.72309$ and $\lambda_{\rm theor}\approx
0.72319$.  On Fig.~1, we compare analytical (third-order expansion) and
numerical results for the invariant distribution.

\begin{figure}[ht]
  \narrowtext \epsfxsize=0.9\hsize \epsfbox{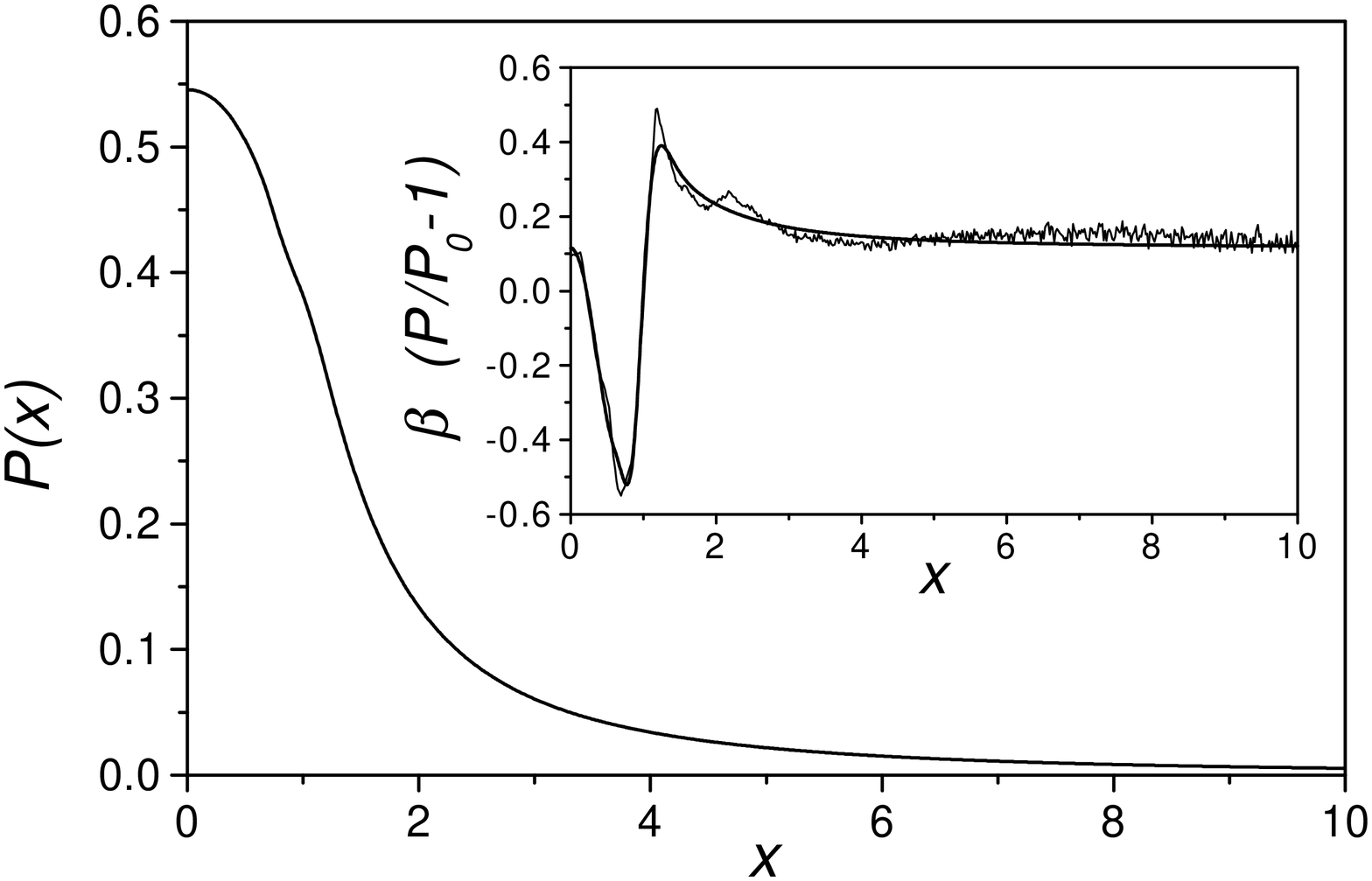}
\caption{$P_{\rm num}$, which is obtained after $2\times 10^9$ iterations
  of the map (\ref{Rnbis}), and $P_{\rm theor}$, which is the exact third
  order expansion in $\beta$, are plotted for $\beta=10$ (curves are
  indistinguishable).  The inset compares $\beta(P_{\rm num}/P_0-1)$ to
  $\beta(P_{\rm theor}/P_0-1)$ (thick line).  The Lyapunov exponent is
  $\lambda_{\rm num}\approx \lambda_{\rm theor}\approx 1.16293$.}
\end{figure}

\section{The Golden Mean case: $\beta=1$}

\subsection{Generalities}

We now focus on the particular case $\beta=1$ which admits a non-perturbative
treatment.  An ingenious construction of the invariant distribution $P(r)$
and the invariant measure $\nu([a,b])=\int_a^b dr\,P(r)$ which involves a
Stern-Brocot division of the real line has been proposed by
Viswanath\cite{V}.  In this subsection we first recall that
construction\cite{V} and the definition of the Stern-Brocot division (for
details, see Ref.\cite{knuth} and also Ref.\cite{conway} which describes
closely related Farey series).  We then derive useful symmetry relations
which will lead to new quantitative results concerning the invariant
distribution $P$ and the invariant measure $\nu$.

The invariant distribution $P(r)$ is symmetric, so we can limit
ourselves to the half-line $r\geq 0$.  The Stern-Brocot division
is defined as follows: Start with the half-line $I_0=[{0\over 1},
{1\over 0}]$ and divide it into two intervals, $[{0\over 1},
{1\over 1}]$ and $[{1\over 1}, {1\over 0}]$; then divide every
first generation interval and continue in this manner. Generally,
the mediant of an interval $[p/q, p'/q']$ is obtained by ``Farey
addition'':
\begin{equation}
\label{fareyadd}
\frac{p}{q}\oplus \frac{p'}{q'}=\frac{p+p'}{q+q'}.
\end{equation}
Thus, the interval $[p/q, p'/q']$ is divided according to the rule
\begin{equation}
\left[\frac{p}{q},\frac{p'}{q'}\right]\longrightarrow
\left[\frac{p}{q},\frac{p+p'}{q+q'}\right]\,\bigcup\,
\left[\frac{p+p'}{q+q'},\frac{p'}{q'}\right].
\label{farey}
\end{equation}
This generates $2^n$ intervals at the $n^{\rm th}$ generation, each interval
$I$ of the $(n-1)^{\rm th}$ generation producing a left and a right child
(${\cal L}I$ and ${\cal R}I$) in the $n^{\rm th}$ generation, according to
Eq.~(\ref{farey}).  For instance, $[{1\over 3}, {1\over 2}]={\cal R}{\cal
  L}{\cal L}I_0$, and generally every Stern-Brocot interval $I$ can be
presented as $I={\cal S}I_0$, where ${\cal S}$ is the unique sequence of
${\cal L}$'s and ${\cal R}$'s.  The length of ${\cal S}$ is denoted $|{\cal
  S}|$.  Using this representation one can prove numerous properties of the
Stern-Brocot division, e.g., the assertion that for every Stern-Brocot
interval $[{p\over q}, {p'\over q'}]$, the identity $p'q-pq'=1$
holds\cite{knuth}.

The invariance condition, i.e. Eq.~(\ref{Prdif}) with $\delta=1$, can be
re-written in terms of the invariant measure to give
\begin{eqnarray*}
\nu(a,b)={1\over 2}\left\{\nu\left({1\over -1+b}, {1\over -1+a}\right)\!+\!
\nu\left({1\over 1+b}, {1\over 1+a}\right)\right\}.
\end{eqnarray*}
Making use of the above equation, it is possible to express the
measure of the left and right children via the measure of the
parent interval\cite{V}:
\begin{equation}
\label{left}
\nu({\cal L}{\cal S}I_0)=
\cases{(1-\tau)\,\nu({\cal S}I_0)  & if $|{\cal S}|$ is even,\cr
       \tau\,\nu({\cal S}I_0)      & if $|{\cal S}|$ is odd,}
\end{equation}
and
\begin{equation}
\label{right}
\nu({\cal R}{\cal S}I_0)=
\cases{\tau\,\nu({\cal S}I_0)      & if $|{\cal S}|$ is even,\cr
      (1-\tau)\,\nu({\cal S}I_0)   & if $|{\cal S}|$ is odd,}
\end{equation}
where $\tau=(\sqrt{5}-1)/2$ is the golden ratio (which is also the inverse
growth constant of the deterministic Fibonacci numbers, $F_n\sim \tau^{-n}$).

We now turn to the derivation of symmetry relations which will be helpful
later.  To obtain the first one we take an arbitrary Stern-Brocot interval
${\cal S}I_0=[{p\over q}, {p'\over q'}]$ and notice that the Stern-Brocot
interval ${\cal SRR}I_0=[{p\over q}+2,{p'\over q'}+2]$ differs from ${\cal
  S}I_0$ by a mere translation.  Using Eqs.(\ref{left})--(\ref{right}) one
finds $\tau(1-\tau)\,\nu({\cal S}I_0)=\nu({\cal SRR}I_0)$, i.e.,
\begin{equation}
\nu\,\left(\left[\frac{p}{q}+2,\frac{p'}{q'}+2\right]\right)=
\rho\,\nu\left(\left[\frac{p}{q},\frac{p'}{q'}\right]\right),
\label{transl}
\end{equation}
with $\rho=\tau(1-\tau)=\sqrt{5}-2$.

Note that every interval with rational end points can be formed by joining
appropriate Stern-Brocot intervals.  Hence Eq.~(\ref{transl}) holds for any
rational interval $[{p\over q}, {p'\over q'}]$.  Since rational numbers form
a dense set on the line and $\nu(r)\equiv \nu([0,r])$ is a continuous
(increasing) function, Eq.~(\ref{transl}) applies to arbitrary intervals.
Specializing to the interval $[0,r]$, we arrive at the first relation
\begin{equation}
\nu\left([2,r+2]\right)=\rho\,\nu(r),
\label{sym1nu}
\end{equation}
which can be also presented as
\begin{equation}
\label{sym1}
P(r+2)=\rho\, P(r) \quad{\rm for}\quad r\geq 0.
\end{equation}

The second symmetry relation,
\begin{equation}
P(2-r)=P(r), \quad{\rm for}\quad 0\leq r\leq 2,
\label{sym2}
\end{equation}
expresses a mirror symmetry with respect to 1 in the interval $[0,2]$.  To
prove this result, we take a Stern-Brocot interval ${\cal SL}I_0=[p/q,p'/q']
\subset [0,1]$ and notice that its symmetric image $[2-p'/q',2-p/q]\subset
[1,2]$ is also a Stern-Brocot interval.  Specifically, the symmetric interval
can be presented as $\bar{\cal S}{\cal LR}I_0$, where $\bar{\cal S}$ is the
sequence obtained from ${\cal S}$ by exchanging the ${\cal L}$'s and ${\cal
  R}$'s.  It turns out that the interval and its symmetric image have the
same invariant measures
\begin{equation}
\nu\left({\cal SL}I_0\right)=\nu\left(\bar{\cal S}{\cal LR}I_0\right).
\label{sym2nu}
\end{equation}
One then deduces Eq.~(\ref{sym2}) from Eq.~(\ref{sym2nu}) repeating the
argument which has been employed in deducing Eq.~(\ref{sym1}) from
Eq.~(\ref{transl}).

The proof of Eq.~(\ref{sym2nu}) can be accomplished by induction on the
length.  Suppose that Eq.~(\ref{sym2nu}) holds for sequences ${\cal S}$ of
length $n$.  Assuming $n$ even and taking the right child of ${\cal SL}I_0$,
we use Eq.~(\ref{right}) to yield
\begin{eqnarray}
\nu\left({\cal RSL}I_0\right)&=&(1-\tau)\, \nu\left({\cal SL}I_0\right),\\
\nu\left({\cal L}\bar{\cal S}{\cal LR}I_0\right)&=&(1-\tau)\,
\nu\left(\bar{\cal S}{\cal LR}I_0\right).
\label{sym2proof}
\end{eqnarray}
Since ${\cal L}=\bar{\cal R}$ and right-hand sides of above equations are
equal, we obtain $\nu\left({\cal RSL}I_0\right)=\nu\left(\bar{\cal R}
\bar{\cal S}{\cal LR}I_0\right)$, thus proving induction step for even $n$
and the right child. The three other cases can be proved in a similar way.

\subsection{Analytical results}

In this subsection, we will make use of the above symmetries to obtain
quantitative results for the invariant distribution $P$ and the invariant
measure $\nu$.  For instance, one can readily establish extreme behaviors.
{}From Eqs.~(\ref{left})--(\ref{right}), one finds
\begin{equation}
\label{nun}
\nu({\cal L}^{2n}I_0)=\nu({\cal R}^{2n}I_0)=\rho^n\,\nu(I_0)
\end{equation}
for any (positive) integer $n$. Clearly, ${\cal L}^{2n}I_0=[0,{1\over 2n}]$
and ${\cal R}^{2n}I_0=[2n,\infty]$.  Additionally, $\nu(I_0)=1/2$ which
immediately follows from normalization.  Thus Eq.~(\ref{nun}) reduces to
\begin{equation}
\label{nuN}
\nu\left({1\over 2n}\right)=\nu([2n,\infty])={1\over 2}\,\rho^n.
\end{equation}
{}From Eq.~(\ref{nuN}) we deduce the small $r$ behavior,
\begin{equation}
\label{small}
\nu(r) \sim e^{-c/r} \quad {\rm with} \quad c=-{1\over 2}\,\ln\rho,
\end{equation}
and the large $r$ behavior,
\begin{equation}
\label{large}
{1\over 2}-\nu(r) \sim e^{-cr}.
\end{equation}

Thus, the invariant measure $\nu(r)$ sharply vanishes when $r\to 0$.
Remarkably, similar plateaux exist around all rational $r$'s.  It is useful
to demonstrate the existence of the plateau at the proximity of a simple
rational point, say $r=1/2$.  Take the family of intervals ${\cal
  L}^{2n}{\cal R}{\cal L}I_0=[{1\over 2},{2n+1\over 4n+1}]$. The length of
the $n^{\rm th}$ interval is $\epsilon_n=[2(4n+1)]^{-1}$.  Exploiting
Eqs.~(\ref{left})--(\ref{right}), we can compute the measure of these
intervals and then examine how the measure scales with length.  We get
\begin{equation}
\label{nu1/2}
\nu\left(\left[{1\over 2},{2n+1\over 4n+1}\right]\right)
={(1-\tau)^2\over 2}\,\rho^n
\sim e^{-c/4\epsilon_n}.
\end{equation}
The continuity of the measure implies that the scaling holds for all small
intervals, $\nu({1\over 2}+\epsilon)-\nu({1\over 2})\sim e^{-c/4\epsilon}$.

Similarly, one can work out the proximity of an arbitrary rational number.
One finds that the measure of the interval $[p/q,p/q+\epsilon]$ ($p$ and $q$
are mutually prime integers) exhibits the following behavior
\begin{equation}
\nu\left(\frac{p}{q}+\epsilon\right)-\nu\left(\frac{p}{q}\right)\sim
\exp\left(-\frac{c}{q^2 \epsilon}\right),
\label{plateau}
\end{equation}
with $c=-{1\over 2}\,\ln\rho$ as above.  Similar asymptotics apply when we
approach the rational point from the left.

Turn now to the invariant distribution $P(r)$.  This function obeys a
striking number of intricate identities. [Note that it is a slight abuse of
language to speak of $P$ as a function: $P$ is a distribution rather than a
function.]  In the following, we will be using Eq.~(\ref{Prdif}), with
$\delta=1$, so we write it down for easy reference
\begin{equation}
\label{Prdif1}
2P(r+1)=\frac{1}{r^2}P\left(\frac{1}{r}\right)+
           \frac{1}{(r+2)^2}\,P\left(\frac{1}{r+2}\right).
\end{equation}
Note that Eq.~(\ref{Prdif1}) at $r+2n$ can be written as
\begin{equation}
\label{Un}
2P(r+2n+1)=U_n(r)+U_{n+1}(r),
\end{equation}
where we used the shorthand notation
\begin{equation}
\label{U}
U_n(r)=\frac{1}{(r+2n)^2}\,P\left(\frac{1}{r+2n}\right).
\end{equation}
Changing $n$ to $n+1$ in Eq.~(\ref{Un}) gives
\begin{equation}
\label{Un1}
2P(r+2n+3)=U_{n+1}(r)+U_{n+2}(r).
\end{equation}
Recalling that $P(r+2n+3)-\rho P(r+2n+1)=0$ [this is Eq.~(\ref{sym1}) at
$r+2n$], we reduce (\ref{Un}), (\ref{Un1}) to a recursion which involves only
$U$'s:
\begin{equation}
\label{Urec}
U_{n+2}(r)+(1-\rho)U_{n+1}(r)-\rho U_{n}(r)=0.
\end{equation}
Since the variable $r$ is mute, Eq.~(\ref{Urec}) is merely a linear recursion
which is straightforwardly solved to find two independent solutions, $(-1)^n$
and $\rho^n$.  Therefore, the general solution is
\begin{equation}
\label{Usol}
\frac{1}{(r+2n)^2}\,P\left(\frac{1}{r+2n}\right)=A(r)(-1)^n+B(r)\rho^n.
\end{equation}
In the $n\to\infty$ limit, the left-hand side of Eq.~(\ref{Usol}) approaches
to zero.  Thus, $A(r)=0$.  Another coefficient $B(r)$ is found by
specializing Eq.~(\ref{Usol}) to $n=0$. Equation (\ref{Usol}) therefore
reduces to
\begin{equation}
\frac{1}{(r+2n)^2}\,P\left(\frac{1}{r+2n}\right)=
\frac{\rho^n}{r^2}P\left(\frac{1}{r}\right).
\label{ex31}
\end{equation}
One can derive a few more useful formulae relating $P(r)$ at different
points. Performing the change of variable $r\to r^{-1}$, one transforms
Eq.~(\ref{ex31}) into
\begin{equation}
P(r)=\frac{\rho^{-n}}{(1+2nr)^2}\, P\left(\frac{r}{1+2nr}\right).
\label{ex3}
\end{equation}
One can also take Eq.~(\ref{ex31}) at $n=1$ and insert it into
Eq.~(\ref{Prdif1}).  The outcome reads
\begin{eqnarray}
\label{Prdif2}
P(r+1)={1+\rho\over 2r^2}\,\,P\left(\frac{1}{r}\right)
\end{eqnarray}
Take now Eq.~(\ref{Prdif2}), change the variable $r\to r+1$, and use
Eq.~(\ref{sym1}). This transforms Eq.~(\ref{Prdif2}) into
\begin{eqnarray}
\label{ex1}
P(r)=\left(\frac{1+\tau}{1+r}\right)^2\, P\left(\frac{1}{1+r}\right).
\end{eqnarray}
We can take the same identity with $(1+r)^{-1}$ instead of $r$ and insert it
into the right-hand side of Eq.~(\ref{ex1}) thus obtaining another identity.
Proceeding in this manner, we arrive at a series of identities
\begin{eqnarray}
\label{exm}
P(r)=\frac{(1+\tau)^{2m}}{(F_{m}+rF_{m-1})^2}\,
P\left(\frac{F_{m-1}+rF_{m-2}}{F_{m}+rF_{m-1}}\right).
\end{eqnarray}
These identities apply for all integer $m$'s as well as the Fibonacci numbers
$F_m$ which are defined for all integer $m$, e.g., $F_{-4}=2, F_{-3}=-1,
F_{-2}=1, F_{-1}=0,F_0=1$.

Replacing $r\to 2-r$ in identities (\ref{exm}) and using the symmetry
relation (\ref{sym2}) one obtains another infinite series of identities.  The
simplest such identity (corresponding to $m=1$) reads
\begin{equation}
P(r)=\left(\frac{1+\tau}{3-r}\right)^2
P\left(\frac{1}{3-r}\right).
\label{ex2}
\end{equation}
The above identities together with the basic relation $\nu([a,b])=\int_a^b
dr\,P(r)$ can be used to re-derive all previous results about the invariant
measure, including the most interesting findings about the plateaux.

\subsection{Generating the invariant measure}

Because of symmetries, it is sufficient to study $\nu(r)$ on the interval
[0,1]. To distinguish the restricted version from the general case we change
the notation: $r\to x$, $P\to {1-\rho\over 4}\,f$, $\nu\to {1-\rho\over
  4}\,N$.  The prefactor $(1-\rho)/4$ guarantees that the invariant measure
$N(x)=\int_0^x dy\,f(y)$ obeys $N(1)=1$. To see this, we take the
normalization condition, $1=\int_{-\infty}^\infty dr P(r)$, and massage it a
bit:
\begin{eqnarray*}
2\int_0^\infty dr\,P(r)={2\over 1-\rho}\int_0^2 dr\,P(r)
={4\over 1-\rho}\int_0^1 dr\,P(r).
\end{eqnarray*}
The first above identity is obtained by using Eq.~(\ref{sym1}) and summing up
the geometric series while the last is an obvious consequence of
Eq.~(\ref{sym2}).  Thus we indeed obtain $N(1)=1$ if we choose $f(x)={4\over
  1-\rho}\,P(x)$.

Now let us integrate $f(x)F(x)$ with any $F(x)$ over intervals $(0,1/3),
(1/3,1/2), (1/2,1)$.  We find
\begin{eqnarray}
\label{F1}
\int_{0}^{1/3} dx\,f\,F &=&\int_0^1\frac{dx}{(1+2x)^2}\,
f\left(\frac{x}{1+2x}\right)\,F\left(\frac{x}{1+2x}\right)\nonumber\\
&=&\rho\int_0^{1}dx\,f(x)\,F\left(\frac{x}{1+2x}\right),\\
\label{F2}
\int_{1/3}^{1/2} dx\,f\,F &=&\int_0^1\frac{dx}{(3-x)^2}\,
f\left(\frac{1}{3-x}\right)\,F\left(\frac{1}{3-x}\right)\nonumber\\
&=&(1+\tau)^{-2}\int_0^{1}dx\,f(x)\,F\left(\frac{1}{3-x}\right),\\
\label{F3}
\int_{1/2}^1 dx\,f\,F &=&\int_0^1\frac{dx}{(1+x)^2}\,
f\left(\frac{1}{1+x}\right)\,F\left(\frac{1}{1+x}\right)\nonumber\\
&=&(1+\tau)^{-2}\int_0^{1}dx\,f(x)\,F\left(\frac{1}{1+x}\right).
\end{eqnarray}
In deriving the first lines in above formulae we have used the mappings
$x(1+2x)^{-1}, (3-x)^{-1}, (1+x)^{-1}$ of the unit interval $(0,1)$ onto the
intervals which appear on the left-hand side.  We then exploited
Eq.~(\ref{ex3}) at $n=1$, Eq.~(\ref{ex2}), and Eq.~(\ref{ex1}), respectively,
to obtain the second lines.

Summing up Eqs.~(\ref{F1})--(\ref{F3}) leads to identity
\begin{eqnarray}
\label{f1}
\int_0^1dx\, f(x)\,F(x)=\int_0^1dx\, f(x)\, \hat T\left[F(x)\right],
\end{eqnarray}
where the linear operator $\hat T$ acts on $F$ according to
\begin{eqnarray}
&&\hat T\left[F(x)\right]
=\sum_{i=1}^3\alpha_i F\left(h_i(x)\right),\label{f2}\\
&& \alpha_1=\rho, \quad \alpha_2=\alpha_3=(1+\tau)^{-2},\quad
\sum_{i=1}^3\alpha_i=1,\label{f3}\\
&&h_1(x)=\frac{x}{1+2x},\quad h_2(x)=\frac{1}{3-x},\quad
h_3(x)=\frac{1}{1+x}.\label{f4}
\end{eqnarray}
These relations show that the distribution $f$ can be generated by the
following simple recursion process.  We take an arbitrary point $x\in (0,1]$
and assign a unit weight $w(x)=1$.  We then define three new points and
associated weights according to the rule
\begin{eqnarray}
x_i=h_i(x),\quad w(x_i)=\alpha_i w(x),\quad i=1,2,3.
\end{eqnarray}
Iterating $n$ times, we get $3^n$ different $x_i$'s.  The weights $w(x_i)$ of
these points add up to 1 and $w(x)$ converge (in the weak sense) to $f(x)$.
In Fig.~2, we plot the invariant measure obtained after 4 and 10 iterations.

\begin{figure}[ht]
  \narrowtext \epsfxsize=0.9\hsize \epsfbox{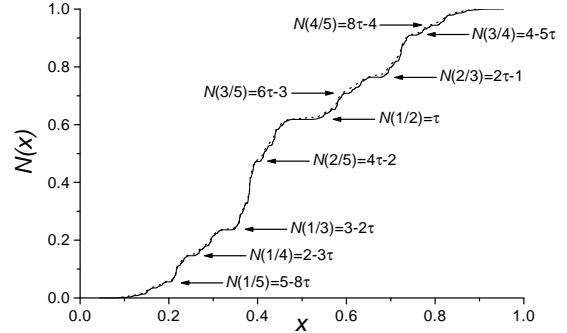}
\caption{The invariant measure $N(x)=\int_0^x dy\,f(y)$ obtained
  after 4 iterations (dotted line involving only $3^4=81$ different $x_i$'s)
  and 10 iterations ($3^{10}=59049$ different $x_i$'s).}
\end{figure}

The plateaux described by Eq.~(\ref{plateau}) are clearly visible
in Fig.~2. The smaller the denominator $q$, the wider is the
associated plateau as predicted by Eq.~(\ref{plateau}). In a
pseudo-gap of $f$ (associated to a plateau for $N$),
$N(p/q)=\{n_{p/q}\tau\}$, where $\{.\}$ denotes the positive
fractional part and $n_{p/q}$ is a possibly negative integer.  The
values of $N(p/q)$ for a few small denominator fractions are presented in
Fig.~2; they have been calculated by using
Eqs.~(\ref{f1})--(\ref{f4}). Thus, a kind of ``pseudo-gap labeling
theorem''\cite{bell} applies here: for tight-binding Hamiltonians
on 1D quasiperiodic chains associated to an irrational number
$\alpha$ (see Ref.\cite{bell} for a general formulation), this
theorem states that the integrated density of
states (IDOS) displays plateaux (corresponding to the energy
true gaps). In these plateaux, the IDOS takes values precisely of
the form $\{n\alpha\}$. In one dimension, the IDOS is intimately related to
the invariant measure $\nu$\cite{JML}. Therefore, it is
not surprising that $\nu$ exhibits a similar behavior.

We now show how to calculate the Lyapunov exponent from the restricted
invariant distribution $f(x)$.  Take the basic formula, $\lambda=\int
dr\,P(r)\,\ln r$, and perform the same massage as we did for the
normalization condition in the beginning of this subsection.  We obtain
\begin{eqnarray*}
&&\lambda=\int_0^1 dx\, f(x)\,g(x)+\mu,\\
&&g(x)=\frac{1-\rho}{2}\sum_{k=0}^\infty \rho^k
\ln\left(1-\left(\frac{1-x}{2k+1}\right)^2\right),\\
&&\mu=(1-\rho)\sum_{k=0}^{\infty}\rho^k\ln(2k+1)=0.29320491137\ldots
\end{eqnarray*}
We then numerically compute $\lambda$ via
\begin{eqnarray*}
\lambda=\lim_{n\to\infty}\sum_{i=1}^{3^n}w(x_i)g(x_i)+\mu.
\end{eqnarray*}
For instance, starting from $x=1-\tau$, one reproduces 10 digits of $\lambda$
after only 8 iterations.

\section{Gaussian Random Sequences}

Invariant measures exhibited by the Fibonacci random sequences appear fractal
for all strengths of disorder.  This is apparently caused by the discreteness
of the disorder.  Additionally, there is a transition between weak disorder,
manifested by an invariant measure whose support is bounded, and strong
disorder characterized by an invariant measure with unbounded support. This
feature is seemingly caused by the boundness of the disorder.  Hence, random
sequences with unbounded continuous disorder distributions provide a natural
counterpart to random sequences with binary disorder.  As a specific example,
we consider the Gaussian random recurrence,
\begin{equation}
\label{rGr}
x_{n+1}=x_n+\beta z_n x_{n-1},
\end{equation}
where $z_n$'s are independent identically distributed random variables with
Gaussian probability density
\begin{equation}
\label{G}
G(z)={1\over\sqrt{2\pi}}\,e^{-z^2/2}
\end{equation}

To analyze the Gaussian random recurrence, we again use the Ricatti variable
$R_n=x_{n+1}/x_n$ which reduces Eq.~(\ref{rGr}) to the random map
\begin{equation}
\label{RnG}
R_n=1+{\beta z_n\over R_{n-1}}.
\end{equation}
The invariant distribution $P(R)$ satisfies
\begin{equation}
\label{PRG0}
P(R)=\int dz\,G(z)
\int dR'\,P(R')\delta\left(R-1-{\beta z\over R'}\right),
\end{equation}
which can be re-written as
\begin{equation}
\label{PRG}
P(R+1)={1\over\sqrt{2\pi\beta^2}}\int dR'\,|R'|\,P(R')\,
e^{-R'^2R^2/2\beta^2}.
\end{equation}
Note two general features of the invariant distribution which directly
follow from Eq.~(\ref{PRG}).  One is a power-law large $R$ asymptotics,
\begin{equation}
\label{Rlarge}
P(R)\to \sqrt{2\over \pi}\,{\beta P(0)\over R^2}
\quad {\rm for}\quad R\gg\beta.
\end{equation}
Another unexpected feature is a weak logarithmic divergence at $R=1$:
\begin{equation}
\label{Rsmall}
P(R)\to {2\over \pi}\,P(0)\,\ln\left({1\over |R-1|}\right)
\quad {\rm for}\quad |R-1|\ll \beta.
\end{equation}
Equation (\ref{Rsmall}) follows from Eqs.~(\ref{PRG})--(\ref{Rlarge}).

We now turn to a perturbative analysis.  While for the Gaussian random
recurrence, there appears to be no threshold separating weak and strong
disorder, different approaches should be implemented when $\beta\to 0$ and
$\beta\to\infty$, respectively. In the former region, the regular
perturbation theory applies while in the latter region one needs a singular
perturbation theory.

\subsection{Weak disorder}

Equation (\ref{RnG}) shows that for weak disorder ($\beta\to 0$), the
magnitude of $R-1$ is comparable with $\beta$.  This suggests to rescale the
variable $R$ and the invariant distribution $P(R)$ according to
\begin{equation}
\label{rQ}
R=1+\beta r, \quad P(R)=\beta^{-1}Q(r).
\end{equation}
The normalization condition $\int dR\,P(R)=1$ now reads
\begin{equation}
\label{norm}
\int dr\,Q(r)=1.
\end{equation}
The governing equation (\ref{PRG}) becomes
\begin{equation}
\label{rQG}
{Q(r)\over G(r)}=\int_{-\infty}^\infty dr'\,|1+\beta r'|\,Q(r')\,
e^{-r^2(\beta r'+\beta^2 r'^2/2)}.
\end{equation}
In the small $\beta$ limit, we simplify Eq.~(\ref{rQG}) to
\begin{equation}
\label{QG}
{Q(r)\over G(r)}=\int_{-\infty}^\infty dr'\,(1+\beta r')\,Q(r')\,
e^{-r^2(\beta r'+\beta^2 r'^2/2)}.
\end{equation}
The error caused by the above simplification is of order $Q(-1/\beta)$.  We
will see that it vanishes as $\exp(-1/2\beta^2)$ and therefore it can be
ignored in perturbative analysis.

We are seeking a perturbative solution.  The symmetry $\beta\leftrightarrow
-\beta$ suggests an expansion in $\beta^2$ rather than $\beta$:
\begin{equation}
\label{Q}
Q(r)=\sum_{n=0}^\infty \beta^{2n} Q_n(r).
\end{equation}
We must also expand the exponent on the right-hand side of Eq.~(\ref{rQG}).
This exponent is the generating function of the Hermite polynomials $H_n(r)$,
\begin{equation}
\label{hermite}
\exp[-r^2(x+x^2/2)]=\sum_{n=0}^\infty {(-xr)^{n}\over n!}\,H_n(r),
\end{equation}
where $H_n(r)$ are defined via\cite{BO}
\begin{equation}
\label{her}
H_n(r)=e^{r^2/2}\left(-{d\over dr}\right)^n e^{-r^2/2},
\end{equation}
Thus, the Hermite polynomials make their appearance in the problem with
non-Hermitian Hamiltonian.

Inserting Eqs.~(\ref{Q})--(\ref{hermite}) into Eq.~(\ref{QG}), we obtain
\begin{eqnarray}
\label{Qn}
{Q_n(r)\over G(r)}=\sum_{k=1}^n {r^{2k-1}\over (2k)!}\,H_{2k+1}(r)
\int dr'\,(r')^{2k}\,Q_{n-k}(r').
\end{eqnarray}
In deriving Eq.~(\ref{Qn}), we also used the recursion relation for the
Hermite polynomials\cite{BO}:
\begin{equation}
\label{herrec}
H_{n+1}(r)=rH_n(r)-nH_{n-1}(r).
\end{equation}
Solving Eq.~(\ref{Qn}) recursively yields
\begin{eqnarray*}
Q_0(r)&=&G(r),\\
{Q_1(r)\over G(r)}&=&{1\over 2}\,rH_3(r),\\
{Q_2(r)\over G(r)}&=&{3\over 2}\,rH_3(r)+{1\over 8}\,r^3H_5(r),\\
{Q_3(r)\over G(r)}&=&12\,rH_3(r)+{5\over 4}\,r^3H_5(r)
+{1\over 48}\,r^5H_7(r),\\
{Q_4(r)\over G(r)}&=&{7857\over 32}\,rH_3(r)+{135\over 8}\,r^3H_5(r)\\
&&+{7\over 16}\,r^5H_7(r)+{1\over 384}\,r^7H_9(r),\\
{Q_5(r)\over G(r)}&=&{1362843\over 256}\,rH_3(r)+{48675\over 128}\,r^3H_5(r)\\
&&+{147\over 16}\,r^5H_7(r)+{3\over 32}\,r^7H_9(r)+{1\over 3840}\,r^9H_{11}(r),
\end{eqnarray*}
etc.  We will use these results to compute the Lyapunov exponent.  The basic
formula (\ref{Lb}) now reads
\begin{equation}
\label{lyap}
\lambda=\int_{-\infty}^\infty dr\,Q(r)\,\ln|1+\beta r|.
\end{equation}
Expanding the logarithm and the invariant distribution, Eq.~(\ref{Q}), and
recalling that $Q(r)=Q(-r)$, we get
\begin{equation}
\label{lexp}
\lambda=-\sum_{n=1}^\infty\beta^{2n} \sum_{k=1}^n {1\over k}
\int_0^\infty dr\,r^{2k}Q_{n-k}(r).
\end{equation}
Inserting above expressions for $Q_n$ ($n=0,\ldots, 5$) into Eq.~(\ref{lexp}),
we obtain the weak disorder expansion:
\begin{eqnarray}
\lambda(\beta)=&-&{1\over 2}\,\beta^2 -{9\over
4}\,\beta^4-22\,\beta^6
-{13197\over 32}\,\beta^8\nonumber\\
&-&{2374335\over 256}\,\beta^{10}-{118392093\over 512}\,\beta^{12}
+{\cal O}(\beta^{14}).
\label{devser}
\end{eqnarray}

Interestingly, neither the $R^{-2}$ asymptotics (\ref{Rlarge}) nor the
logarithmic singularity (\ref{Rsmall}) appear in the weak disorder expansion.
Both these behaviors are non-perturbative.  For instance, $P(0)$ which
appears on the right-hand sides of Eqs.~(\ref{Rlarge})--(\ref{Rsmall}) scales
as $\exp(-1/2\beta^2)$, i.e., the behaviors
Eqs.~(\ref{Rlarge})--(\ref{Rsmall}) are beyond the scope of perturbation
techniques.  Note also that the observation of the logarithmic singularity
(\ref{Rsmall}) requires probing a prohibitively tiny region
$r\sim\exp[-\exp(-1/2\beta^2)]$.

The occurrence of non-perturbative corrections (of order $\exp(-1/2\beta^2)$)
suggests that the radius of convergence of the series (\ref{devser}) is equal
to zero.  This is (non-rigorously) confirmed by the following approximate
analysis of $\lambda(\beta)$.  First, we write $R_n\approx 1+\beta z_n$, as
the distribution of $R_{n-1}$ is strongly peaked at $R=1$ for small $\beta$.
Within this approximation, and using the fact that the Gaussian distribution
is an even function, we obtain
\begin{equation}
\lambda(\beta)\approx\frac{1}{2}\int_{-\infty}^\infty dz\,
G(z)\ln|1-\beta^2z^2|\,dz,
\label{app}
\end{equation}
which reproduces exactly the first term of Eq.~(\ref{devser}).
Now, expanding Eq.~(\ref{app}) in powers of $\beta^2$, is likely
to lead to the correct qualitative behavior for the full general
expansion.  The generic term,
\begin{eqnarray*}
-{\beta^{2n}\over 2n}\int_{-\infty}^\infty dz\, G(z)\,z^{2n}=
-{\beta^{2n}\,2^{n-1}\over n\,\sqrt{\pi}}\Gamma\left(n+\frac{1}{2}\right),
\end{eqnarray*}
grows faster than any exponential, ensuring that the radius of convergence is
indeed zero. Of course, we cannot exclude that for the actual expansion,
subtle cancellations lead to a finite radius of convergence. However, the
occurrence of non-perturbative corrections, and concrete ingredients of the
argument presented above (mainly, the unboundness of the distribution $G(z)$
and the fact that the series $\ln(1+x)$ as a finite radius of convergence)
which seem to persist in the general case, favor a zero radius of convergence
and the asymptotic character of the series (\ref{devser}). This is very
different from the case of random Fibonacci sequences where the weak disorder
expansion has a finite radius of convergence and there were no trace of any
non-perturbative contribution.

\subsection{Strong disorder}

For $\beta\to\infty$, we again use the properly normalized Ricatti variable
$y_n=x_{n+1}/x_n\sqrt{\beta}$.  Equation (\ref{rGr}) reduces to the
random map
\begin{equation}
\label{ynG}
y_n={z_n\over y_{n-1}}+\delta, \qquad
\delta\equiv \beta^{-1/2}.
\end{equation}
The invariant distribution satisfies
\begin{eqnarray}
\label{PyG}
P(y+\delta)={1\over\sqrt{2\pi}}\int_{-\infty}^\infty d\eta\,|\eta|\,P(\eta)\,
\exp\left\{-{y^2 \eta^2\over 2}\right\},
\end{eqnarray}
and the Lyapunov exponent is given by Eq.~(\ref{lb}) as in the Fibonacci
case.  Equation (\ref{PyG}) suggests to seek a perturbative solution.  In the
zeroth order, one might set $\delta=0$ in Eq.~(\ref{PyG}).  The corresponding
invariant distribution $P_0(y)$ is an even function of $y$ which satisfies
\begin{eqnarray}
\label{Py}
P_0(y)=\sqrt{2\over \pi}\int_0^\infty d\eta\,\eta\,P_0(\eta)\,
\exp\left\{-{y^2 \eta^2\over 2}\right\}.
\end{eqnarray}
Paradoxically, a (formal) solution to this equation,
\begin{equation}
\label{formal}
P_0(y)={A\over |y|},
\end{equation}
does not obey the normalization requirement. This indicates that the naive
perturbation approach does not work and one must develop a singular
perturbation theory.  One still anticipates that $P_0(y)$ is given by
Eq.~(\ref{formal}) apart from the small and large scales, $|y|\sim \delta$
and $|y|\sim \delta^{-1}$, which are implied by the random map (\ref{ynG}).
Treating these scales as cutoffs allows us to normalize the solution
(\ref{formal}) and to estimate the amplitude $A\approx (2\ln\beta)^{-1}$.
One can establish the existence of the cutoffs more rigorously.  Using
Eqs.~(\ref{PyG}), (\ref{formal}) one estimates $P(0)\sim A\delta^{-1}$ in
agreement with the existence of the small scale cutoff $y\sim\delta$.  The
large scale cutoff already follows from Eq.~(\ref{Rlarge}) which now reads
\begin{equation}
\label{ylarge}
P(y)\to \sqrt{2\over \pi}\,{P(0)\over y^2}
\quad {\rm for}\quad y\gg\delta^{-1}.
\end{equation}

Note also that the deficiency of the naive perturbation approach is clear
from the respective random map, $y_n=z_n/y_{n-1}$.  Indeed, iterating the
above map and taking the logarithm gives $\ln y_{n}=\sum (-1)^{n-k}\ln z_k$.
The central limit theorem now asserts that the scale of the limiting the
distribution grows indefinitely with $n$, namely $\ln y_n\sim\sqrt{n}$.  Thus
for $\delta=0$ already the basic assumption that $y_n$ approaches to a
limiting distribution which does not depend on $n$ is incorrect.

We now present a computation of the zeroth order contribution to the Lyapunov
exponent which does not require the knowledge of $P_0$.  Denote by $\Lambda$
the zeroth order contribution to the Lyapunov exponent.  We have
\begin{eqnarray*}
\Lambda&=&2\int_0^\infty dy\,P_0(y)\,\ln y\\
&=&2\sqrt{2\over \pi}\int_0^\infty d\eta\,\eta\,P_0(\eta)
\int_0^\infty dy\,\ln y\,\exp\left\{-{y^2 \eta^2\over 2}\right\}\\
&=&2\sqrt{2\over \pi}\int_0^\infty d\eta\,P_0(\eta)
\int_0^\infty {dt\over \sqrt{2t}}\,e^{-t}\,
\ln\left({\sqrt{2t}\over\eta}\right)\\
&=&\int_0^\infty d\eta\,P_0(\eta)
\left[\Psi\left({1\over 2}\right)+\ln 2 - 2\ln\eta\right]\\
&=&{1\over 2}\left[\Psi\left({1\over 2}\right)+\ln 2\right]-\Lambda.
\end{eqnarray*}
In the second line we used Eq.~(\ref{Py}); this step is not really rigorous
though we think the final result is correct.  The variable $t$ which appears
in the third line has been defined via $t=y^2\eta^2/2$; in the fourth line we
used the digamma (psi) function, $\Psi(x)=\Gamma'(x)/\Gamma(x)$\cite{BO}; in
the last line we used the normalization requirement and the definition of
$\Lambda$. The above equation yields $\Lambda$ which can be simplified
further by using the identity $\Psi(1/2)=-\gamma-2\ln 2$, where $\gamma$ is
the Euler constant.  Finally,
\begin{equation}
\label{Lambda} \Lambda=-{\gamma+\ln 2\over
4}=-0.317590711365\ldots
\end{equation}

Our numerical results suggest that the strong disorder expansion involves
powers of $(\ln\beta)^{-1}$ rather than $\beta^{-1}$:
\begin{equation}
\label{Lambda1} \lambda(\beta)={1\over
2}\ln\beta+\Lambda+\sum_{k=1}^\infty b_k(\ln\beta)^{-k}.
\end{equation}
Of course, it is hardly possible to probe higher order
logarithmic terms numerically. However, plotting $\lambda-{1\over
2}\ln\beta$ versus $(\ln\beta)^{-1}$ gives a fairly straight line
for $\beta>10^4$, with the slope $b_1\approx 0.557$, and a perfect
fit to the above functional form, keeping a quadratic term in
$(\ln\beta)^{-1}$, with $b_2\approx -0.52$ (see Fig.~3).

\begin{figure}[ht]
\narrowtext \epsfxsize=0.9\hsize \epsfbox{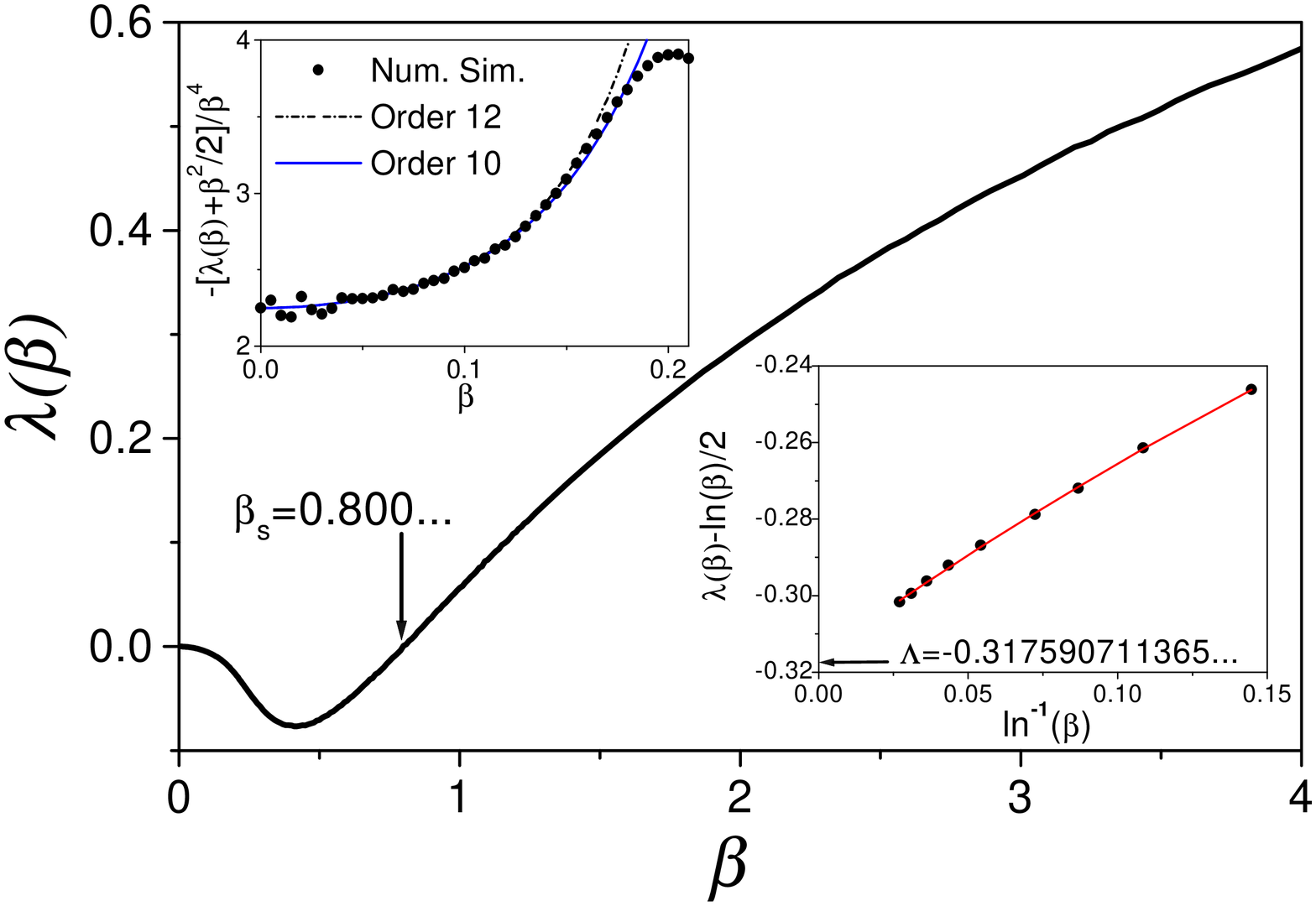}
\caption{We plot $\lambda(\beta)$ for Gaussian random sequences.
  The upper insert compares $-(\lambda(\beta)+\beta^2/2)/\beta^4=9/4
  +22\beta^2+...$ obtained from the 10th and 12th order expansions in
  Eq.~(86) to the result of numerical simulations. The lower insert is a
  quadratic fit in $(\ln\beta)^{-1}$ (see Eq.~(94)) of
  $\lambda(\beta)-{1\over 2}\ln\beta$, for large $\beta$, leading to an
  extrapolated $\Lambda\approx-0.317$, with $b_1\approx 0.557$ and
  $b_2\approx - 0.52$.}
\end{figure}

Extrapolating the quadratic fit to $(\ln\beta)^{-1}=0$ yields
$\Lambda\approx -0.317$, in  good agreement with the theoretical
prediction (\ref{Lambda1}).  Contrary to the Fibonacci
case, the curve of $\lambda(\beta)$ appears perfectly smooth, and
is certainly not fractal.

Finally, we note that asymptotic methods\cite{BO} should in
principle allow to perform the strong disorder expansion more
systematically.

\section{Discussion}

The rich behavior exhibited by random Fibonacci numbers suggests avenues for
further investigation.  For instance, how to reconcile perturbative results
in the large $\beta$ limit with non-perturbative results for $\beta=1\,$?
This question is important as there appears to be just a single threshold
$\beta=1/4$ and therefore $\beta=1$ lies within the large $\beta$ domain.
This suggests qualitative similar behaviors which is not the case.  The major
difference between $\beta=1$ and $\beta\to\infty$ cases is manifested in
extreme behaviors of the invariant measure.  In the former case, it exhibits
exponential asymptotics, Eqs.~(\ref{small})--(\ref{large}), while the latter
is characterized by power-law asymptotics, $\nu(r)\sim r$ for $r\to 0$ and
${1\over 2}-\nu(r) \sim r^{-1}$ for $r\to\infty$.  More generally, our
perturbative results are infinitely smooth, in a gross disagreement with the
behavior for $\beta=1$.  Another (related) set of questions concerns the
curve $\lambda(\beta)$: Is it a fractal? Does it become genuinely smooth at
least for sufficiently large $\beta\,$?

One could ask for a more complete characterization of the growth (decay) of
the random Fibonacci numbers.  A natural conjecture is $x_n\sim e^{\lambda
  n}\,n^\omega$.  For the neutrally stable recurrence, $x_{n+1}=x_n \pm
\beta_s x_{n-1}$ with $\beta_s\approx 0.70258$ chosen so that
$\lambda(\beta_s)=0$, the above conjecture would imply the power-law behavior
$x_n\sim n^{\omega_s}$.

Finally, it would be very interesting to analyze random Fibonacci numbers for
the critical strength of disorder, $\beta=1/4$.  This strength of disorder
appears a bit more interesting than $\beta=1$: It is hard to see what
distinguishes $\beta=1$ from say $\beta=1.23456$, while the case of
$\beta=1/4$ is certainly special as it separates the regions of weak and
strong disorder.

\smallskip
\noindent
We are grateful to H.~Castillo, G.~Oshanin, S.~Redner, and P.~Sharma
for discussions.  The work of PLK was supported by NSF Grant DMR9978902.

\end{multicols}
\end{document}